\documentstyle[12pt]{article}    
\begin{document}
\begin{center}

     {\bf     THERMODYNAMICS OF NON-ABELIAN EXCLUSION STATISTICS } 

\vspace{1cm}

                      Wung-Hong Huang\\
                       Department of Physics\\
                       National Cheng Kung University\\
                       Tainan,70101,Taiwan\\

\end{center}
\vspace{2cm}
~~~The thermodynamic potential of ideal gases described by the simplest
non-abelian statistics is investigated.   I show that the potential is the
linear function of the element of the abelian-part statistics matrix.  
Thus, the factorizable property in the Haldane (abelian) fractional
exclusion shown by the author [W. H. Huang, Phys. Rev. Lett. 81, 2392
(1998)] is now extended to the non-abelian case.   The complete expansion
of the thermodynamic potential is also given.

\vspace{3cm}

\begin{flushleft}

Classification Number: 05.30.-d, 71.10.+x \\
E-mail:  whhwung@mail.ncku.edu.tw\\
     
\end{flushleft}

\newpage

\begin{center}    {\bf  I. INTRODUCTION} \end{center}

      It is know that the quasiparticle appearing in the strong correlation
system in the low dimension may obeying the Haldane fractional exclusion
statistics [1].   The famous examples are those in the fractional quantized
Hall effect and spin 1/2 antiferromagnetic chain [2-4].  The thermodynamics
in these system had been investigated by several authors [5-8].

    Since the braid group in two space dimensions may be represented by the
non-commute matrix, the non-abelian braid statistics could be shown in a
real world [9-16].   These include the quasiparticle in the non-abelian
fractional quantum Hall state as well as those in the conformal field
theory [9-15].   The spinon in the  Heisenberg chains could also show the
non-abelian exchange statistics [16]. 

     In a recent paper [15], Guruswamy and Schoutens had derived the
occupation number distribution functions for the particle obeying the
non-abelian statistics.   They proposed the following equations for $M$
types of particles and $k$ types of pseudo-particles

     $$(\frac{\Lambda_A -1}{\Lambda_A})  \prod_B  \Lambda_B^{\alpha_{AB}} 
\prod_i  \Lambda_i^{G_{Ai}} = z_A  , ~~~~~A = 1, ...M,  \eqno{(1.1a)}$$   
     $$(\frac{\Lambda_i -1}{\Lambda_i})  \prod_A  \Lambda_A^{G_{iA}} 
\prod_j  \Lambda_j^{G_{ij}} = 1 , ~~~~~i = 1, ...k-1,  \eqno{(1.1b)}$$

\noindent 
where $\alpha_{AB}$ is the abelian-part statistics matric, $G_{iA} = G_{Ai}
= - \frac{1}{2} \delta_{i,1}$ and $G_{ij} = \frac{1}{2} (C_{k-1})_{ij}$,
with $C_{ij}$ the Cartan matrix of the associated group.   Note that 
$\Lambda_a$ is the single-level grand canonical partition function and $z_a
\equiv e^{\beta (\mu_a -\epsilon)}$, with $\epsilon$ the energy level and
$\mu_a$ the chemical potential of the particle of type $a$.   When $G_{Ai}
=0 $ Eq.(1.1a) describe the  Haldane fractional exclusion statistics [5]. 

    The Eqs.(1.1a) and (1.1b) are very similar to those in the abelian case
[5].    However, the absence of the $z_i~~(i=1...(k-1))$ in Eq.(1.1b) means
that the $k$ pseudo-particles do not be suppressed at high energy.    This
property is the characteristics of the  non-abelian exclusion statistics
[15]. 

      In this letter I will analyze the case of (M,k)=(1,2).   This is the
simplest extension of the abelian to the non-abelian case.  Despite the
simplicity this case can describe the q-pfaffian non-abelian fractional
quantum Hall state [12,13].   I will follow the method in my previous
letter [7] to investigate the thermodynamic potential $Q_\alpha$.    A
non-perturbative proof to is used to show that  $Q_\alpha$ can be
factorized in terms of these in the, so called,  non-abelian  boson
($\alpha =0$) and in the non-abelian  fermion ($\alpha =1$), i.e.,

	$$Q_\alpha(N) = (1-\alpha)Q_0 (N) + \alpha Q_1(N). \eqno{(1.2)}$$

\noindent
Thus the factorizable property of the thermodynamical potential found in
the Haldane (abelian) fractional exclusion  statistics also appears in the
non-abelian extension.    As I also present the complete expansion of the
thermodynamic potential.

\begin{center}    {\bf  II. FACTORIZATION AND VIRIAL EXPANSION}
\end{center}

    For simplest case  (M,k)=(1,2)  Eqs.(1.1a) and (1.1b) become

   $$(\frac{\Lambda -1}{\Lambda})   \Lambda^{\alpha}  \Lambda_1^{-
\frac{1}{2}} = z , ~~~~~~(\frac{\Lambda_1 -1}{\Lambda_1}) \Lambda ^{-
\frac{1}{2}}  \Lambda_1 = 1 .  \eqno{(2.1)}$$

\noindent
Eliminating $\Lambda_1$ from the above equation we have

$$(\frac{\Lambda -1}{\Lambda})   \Lambda^{\alpha} (1+
\Lambda^{\frac{1}{2}}) ^{-\frac{1}{2}} = z .   \eqno{(2.2)}$$

\noindent
 For later analysis the above equation is  expressed as
$$\ln (\Lambda -1) + (\alpha -1) \ln\Lambda  -\frac{1}{2} \ln (1+
\Lambda^{\frac{1}{2}}) =  \ln z = \beta (\mu -\epsilon) .   \eqno{(2.3)}$$

    Since the distribution function  $n$ is defined by [15]

             $$ n = \frac{d \ln \Lambda}{d \ln z},    \eqno{(2.4)}$$

\noindent
we can from Eq.(2.3) find that

$$ n = \frac {1} {\frac{1}{4} \frac{1}{1+ \Lambda^{\frac{1}{2}}}+ \frac
{1}{ \Lambda -1} + (\alpha - \frac {1}{4})}.   \eqno{(2.5)}$$

\noindent
Using this relation the differentiation of Eq.(2.3) with respect to 
$\alpha$ and $\epsilon$ will lead to two simple relations

        $$\frac{d}{d \alpha}\ln \Lambda  = n ~ (\beta  \frac{d \mu}{d
\alpha} - \ln \Lambda) , \eqno{(2.6)}$$
          $$\frac{d}{d \epsilon}\ln \Lambda  = - n \beta,  \eqno{(2.7)}$$

\noindent
respectively.

  Using the above equation we see that

    $$N \equiv \int _0^\infty   d(\epsilon \beta) \frac{V}{\lambda ^2} n 
            = -   \int _0^\infty   d \epsilon \frac{V}{\lambda ^2}
\frac{d}{d \epsilon}\ln \Lambda  =   \frac{V}{\lambda ^2} \ln \Lambda _0 ,
~~ \Rightarrow  \Lambda _0 = e^{\frac{N \lambda ^2}{V}} \eqno{(2.8)}$$

\noindent
where $N$ is the particle number and $\Lambda _0$  is the zero-energy 
grand canonical  partition function.     (Note that the system we considerd
here has only one type of real particle, but two types of pseudo-particles.
    $N$ denotes the total number of the real particles.)   Now,
substituting the above relation into Eq.(2.3) (letting $\epsilon =0 $) the
exact form of the chemical potential can be found

  $$ \beta \mu = \alpha  \frac{N \lambda ^2}{V}  + \ln (1 - e^{- \frac{N
\lambda ^2}{V}}) - \frac{1}{2} \ln (1 + e^{\frac{1}{2}\frac{N \lambda
^2}{V}}). \eqno{(2.9)}$$

\noindent
Thus the chemical potential $\mu$  is a linear function of the abelian-part
statistics parameter $\alpha$.   Note that the first two term in the above
equation are exactly those in the abelian exclusion statistics [5].   It is
the last $\log$ term which becomes the new contribution in the non-abelian
exclusion statistics.  

   Next, consider the differentiation of the thermodynamic potential
$Q_\alpha$ with respect to $\alpha$

$$ \frac{dQ_\alpha}{d\alpha} = -kT \int _0^\infty   d(\epsilon \beta)
\frac{V}{\lambda ^2} \frac{d}{d\alpha} \ln \Lambda 
  =  -kT \int _0^\infty   d(\epsilon \beta) \frac{V}{\lambda ^2} ~ n~(
\beta  \frac{d \mu}{d \alpha} - \ln \Lambda) $$ 

     $$ =  -kT \int _0^\infty   d(\epsilon \beta) \frac{V}{\lambda ^2} ~ n~
                            \frac{N \lambda ^2}{V} - kT \int _0^\infty  d
\epsilon \frac{V}{\lambda ^2}  \ln \Lambda \frac{d}{d \epsilon}\ln \Lambda 
=  -   \frac {1}{2} NkT  \frac{N \lambda ^2}{V},   \eqno{(2.10)}$$

\noindent
in which the relations Eqs.(2.6)-(2.9) have been used.  Thus the
thermodynamical potential $Q_\alpha$ is a linear function of the
abelian-part statistics parameter $\alpha$.   This means that the potential
can be exprssed as $Q_\alpha = f + \alpha ~g $, in which $f$ and $g$ do not
depend on the $\alpha$.   Then, if  $\alpha = 1$ we find that $Q_1 = f + g
$, and if  $\alpha = 0$ we find that $Q_0 = f$.     Thus $g =  Q_1 - Q_0 $
and $ f = Q_0$ and we have the factorizable property in Eq.(1.2).    

    The remain work is to find the Virial expansion of the potential
$Q_\alpha$.

    To do this we can first use the method of integration by part to
replace the integration of $\epsilon$  by $\ln \Lambda$.   Then, using the
relations Eqs.(2.3), (2.8) and (2.9)  we have

   $$ Q_\alpha = -kT \int _0^\infty   d(\epsilon \beta) \frac{V}{\lambda
^2}  \ln \Lambda = kT \int _{\ln\Lambda_0}^0 d(\ln \Lambda)
\frac{V}{\lambda ^2} \epsilon \beta
~~~~~~~~~~~~~~~~~~~~~~~~~~~~~~~~~~~~~~~~~~~~~~ $$
$$ =  kT \int _{\ln \Lambda_0}^0   d(\ln \Lambda) \frac{V}{\lambda ^2} \{
\beta \mu - [\ln (\Lambda -1) + (\alpha -1) \ln\Lambda  -\frac{1}{2} \ln
(1+ \Lambda^{\frac{1}{2}})] \} ~~~~~~~~~~~~~~~~$$
  $$ =-  kT (\ln \Lambda_0) \frac{V}{\lambda ^2} \beta \mu 
- kT  \int _{\ln \Lambda_0}^0 d(\ln \Lambda) \frac{V}{\lambda ^2}  [\ln
(\Lambda -1) + (\alpha -1) \ln\Lambda  -\frac{1}{2} \ln (1+
\Lambda^{\frac{1}{2}})] $$
$$= -\frac{1}{2} \alpha NkT  \frac{N \lambda ^2}{V}- kT \int_0^{ \frac{N
\lambda ^2}{V}} dx \frac{V}{\lambda ^2} \frac{x}{e^x -1} +\frac{1}{4} kT
\int_0^{ \frac{N \lambda ^2}{V}} dx \frac{V}{\lambda ^2} \frac{x
e^{x/2}}{e^{x/2}+1},~~~~~~~~ \eqno{(2.11)}$$

\noindent
in which the first integration can be expressed as the summation of the
Bernoulli
numbers $B_l(0)$, as that in the abelian case.   The second integration is
the new contribution from non-abelian statistics, which can be expressed as
the summation of the Euler function $E_l(1)$.     Using the mathematic
definition

  $$ \frac{t e^{xt}}{e^t -1} \equiv \sum_{l=0}^\infty  B_l (x)
\frac{t^l}{l!},$$
  $$ \frac{2  e^{xt}}{e^t + 1} \equiv \sum_{l=0}^\infty  E_l (x)
\frac{t^l}{l!}.$$

\noindent
We thus find the final result

   $$ Q_\alpha  =  -\frac{1}{2} \alpha NkT  \frac{N \lambda ^2}{V}- NkT ~
[\sum_{l=0}^\infty B_l(0) \frac {1}{(l+1)!} (\frac{N \lambda ^2}{V})^l -
\frac{1}{4} \sum_{l=0}^\infty E_l(1) \frac {l+1}{(l+2)!} (\frac{N \lambda
^2}{2V})^{l+1}] $$

$$= -\frac{1}{2} \alpha NkT  \frac{N \lambda ^2}{V}- kTN[1 - \frac{5}{16}
\frac{N \lambda ^2}{V} +
\frac{5}{288}(\frac{N \lambda ^2}{V})^2 - \frac{17}{115200}(\frac{N \lambda
^2}{V})^4 ~~~~~~$$
$$+ \frac{13}{5419008}(\frac{N \lambda^2}{V})^6 -
\frac{257}{5573836800}(\frac{N \lambda ^2}{V})^8 + ... ], \eqno{(2.12)}$$

\noindent
which is the Virial expansion to the eighth order and is consistent with
the Eq.(2.10).

     It is seen that, as that in the abelian case [7],  only the second
Virial coefficient depends on the abelian-part Haldane statistics
parameter.   This means that only the zero-temperature quantities could
depend on the Haldane statistics parameter.

\begin{center}    {\bf  III.  CONCLUSION} \end{center}

   In this paper I have investigated the thermodynamics of the simplest
case, (M,k)=(1,2), of the non-abelian exclusion statistics.     Despite its
simplicity the case can be used to described the real world  system.    The
q-pfaffian non-abelian fractional quantum Hall state  is a special case
with $\alpha = \frac{q+1}{4q}$   [12,13,15].   I have shown that the
thermodynamic potential of the ideal gases obeying the simplest non-abelian
exclusion statistics is a linear function of the (only) element of the
abelian-part statistics matix.    The factorizable property found in the
abelian exclusion is thus now extended to the simplest non-abelian case.   
It is interesting to see that, in both of abelian and non-abelian case,
only  zero-temperature quantities could depend on the Haldane statistics
parameter.     

    A little extension is that with  arbitrary M (while remaining k=2).  
In this case  $\Lambda_1= 1+ \prod_A  \Lambda_A^{G_{1A}}$ (form Eq.(1.1b)).
  After substituting this into Eq.(1.1a)  the pseudo-particle part is
eliminated  and thermodynamical potential could be studied with more
analyses [8].   However, in the general system there may be  more then one
pseudo-particle, i.e., if $k > 2$ in Eq.(1.1b).    In this system the
pseudo-particle part, $\Lambda_i$, could not be easily eliminated from the
Eqs.(1.1a) and (1.1b).   Thus the situation is more complex and the
property of the thermodynamic potential is difficult to be analyzed.   
These problems are remained in my furthermore investigation.

\newpage

\begin{enumerate}
\item  F. D. M. Haldane, Phys. Rev. Lett. 67, 937 (1991).
\item   R. B. Laughlin, Phys. Rev. Lett, 50, 1359 (1983); Phys. Rev. B 27,
3383 (1983); I. Halperin, Phys. Rev. Lett. 52, 1583 (1984); 52, 2390(E)
(1984).
\item   R. Prange and S. M. Girvin, The quantum Hall effect (Springer,
Berlin, 1987).
\item   R. B. Laughlin, 60, 2677 (1988); A. L. Fetter, C. B. Hanna, and R.
B. Laughlin, Phys. Rev. B39, 9679 (1989).
\item  Y. S. Wu, Phys. Rev. Lett. 73, 922 (1994).
\item  C. Nayak,and F. Wilczek, Phys. Rev. Lett. 73, 2740 (1994); A. K.
Rajagopal, ibid. 74, 1048 (1995); M. V. N. Murthy and R. Shankar, ibid. 72,
3629 (1994); W. H. Huang, Phys. Rev. E51, 3729 (1995); W. H. Huang, Phys.
Rev. B 53, 15842 (1996).
\item  W. H. Huang, Phys. Rev. Lett. 81, 2392 (1998).
\item   W. H. Huang, Phys. Rev. B 60, 15742 (1999).
\item  G. Moore and N. Read, Nucl. Phys. B 360, 362 (1991). 
\item  X. G. Wen, Phys. Rev. Lett. 66, 802 (1991).
\item  C. Nayak and F. Wilczek, Nucl. Phys. B 479, 529 (1996); E. Fradkin,
C. Nayak and K. Schoutens, B 546, 711 (1999). 
\item   N. Read and E. Rezayi, Phys. Rev. B 54, 16864 (1996); B 59, 8084
(1999).
\item   K. Schoutens, Phys. Rev. Lett. 79, 2608 (1997); R. A. J. van.
Elburg and  K. Schoutens, Phys. Rev. B58,15704 (1998); P. Bouwknegt and K.
Schoutens, Nucl. Phys. B 547, 501 (1999).
\item  P. Bouwknegt, A. W. W. Ludwig and K. Schoutens, Phys. Lett. B 359,
304 (1995).
\item  S. Guruswamy and K. Schoutens, Nucl. Phys. B 556, 530 (1999).
\item   H. Fahm and M. Stahlsmeier, Phys. Lett. A 250, 293 (1998).
\end{enumerate}

\end{document}